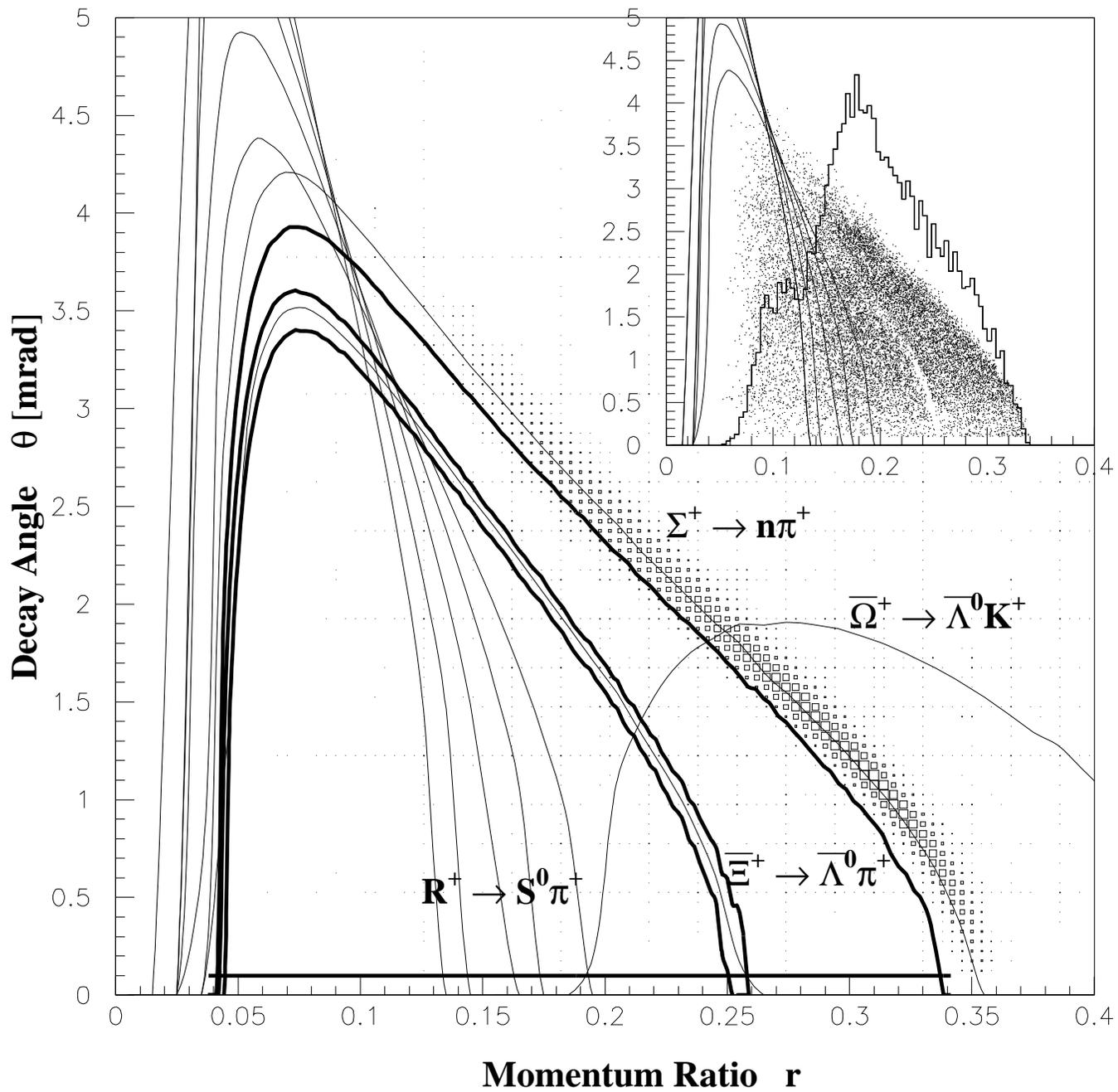

Fig. 1

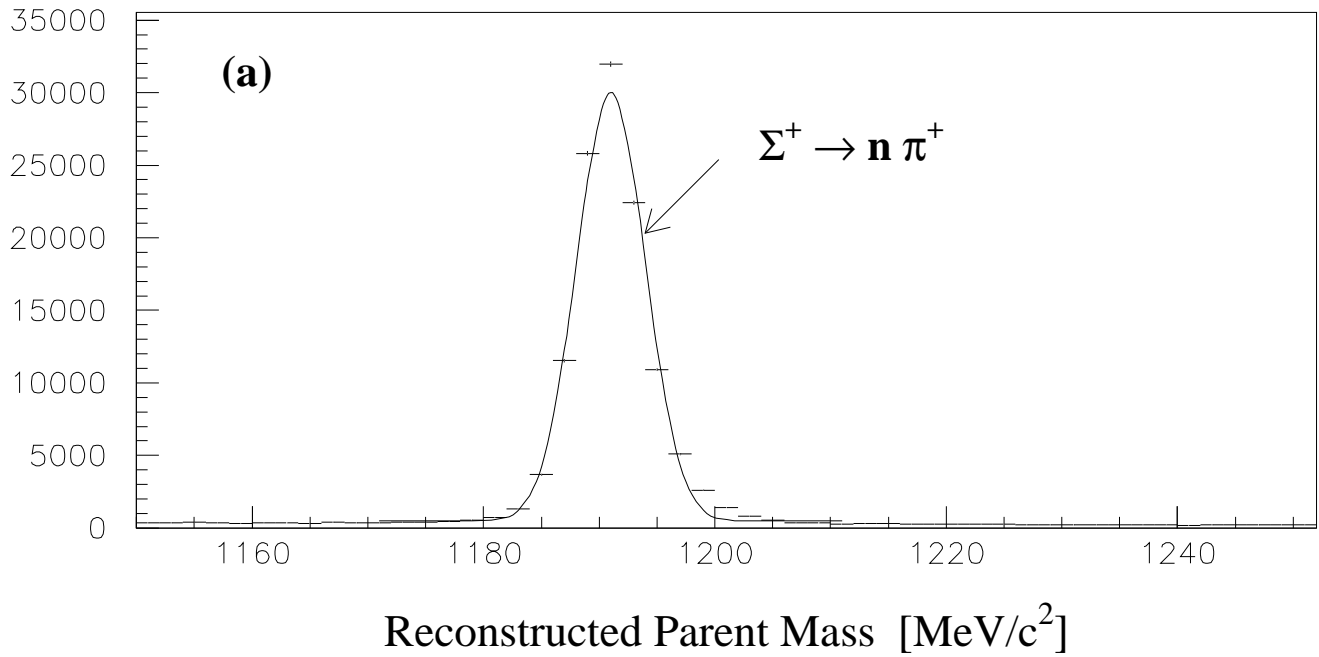
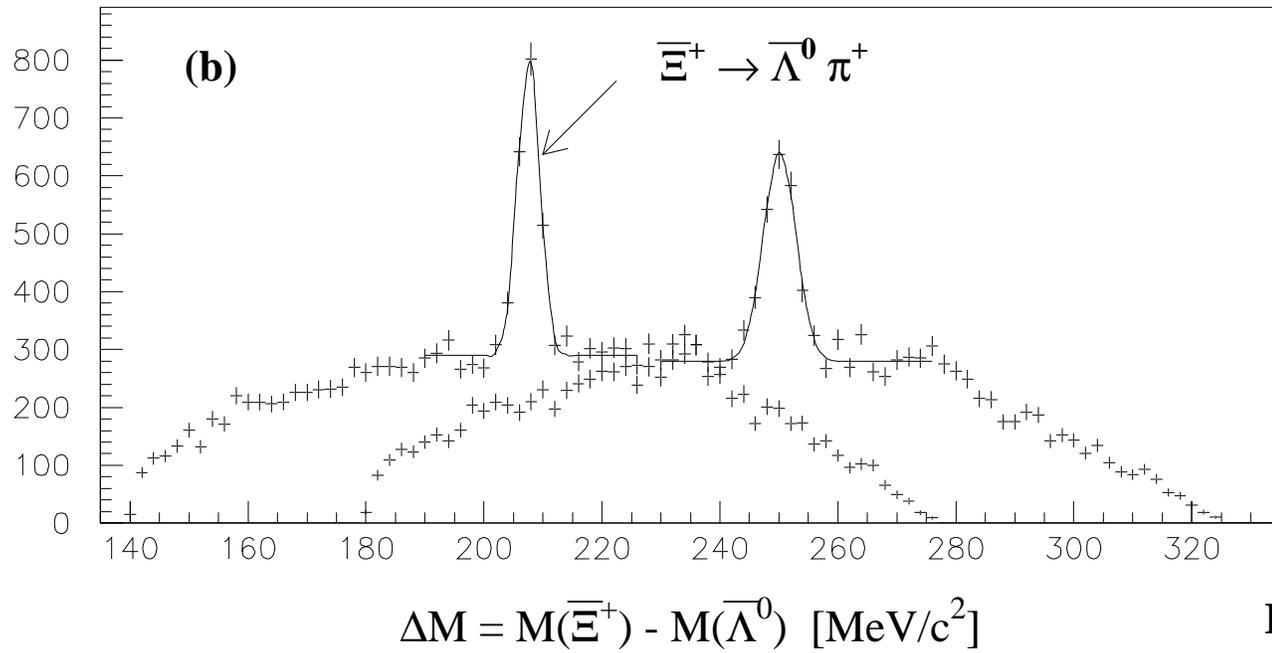

Fig. 2

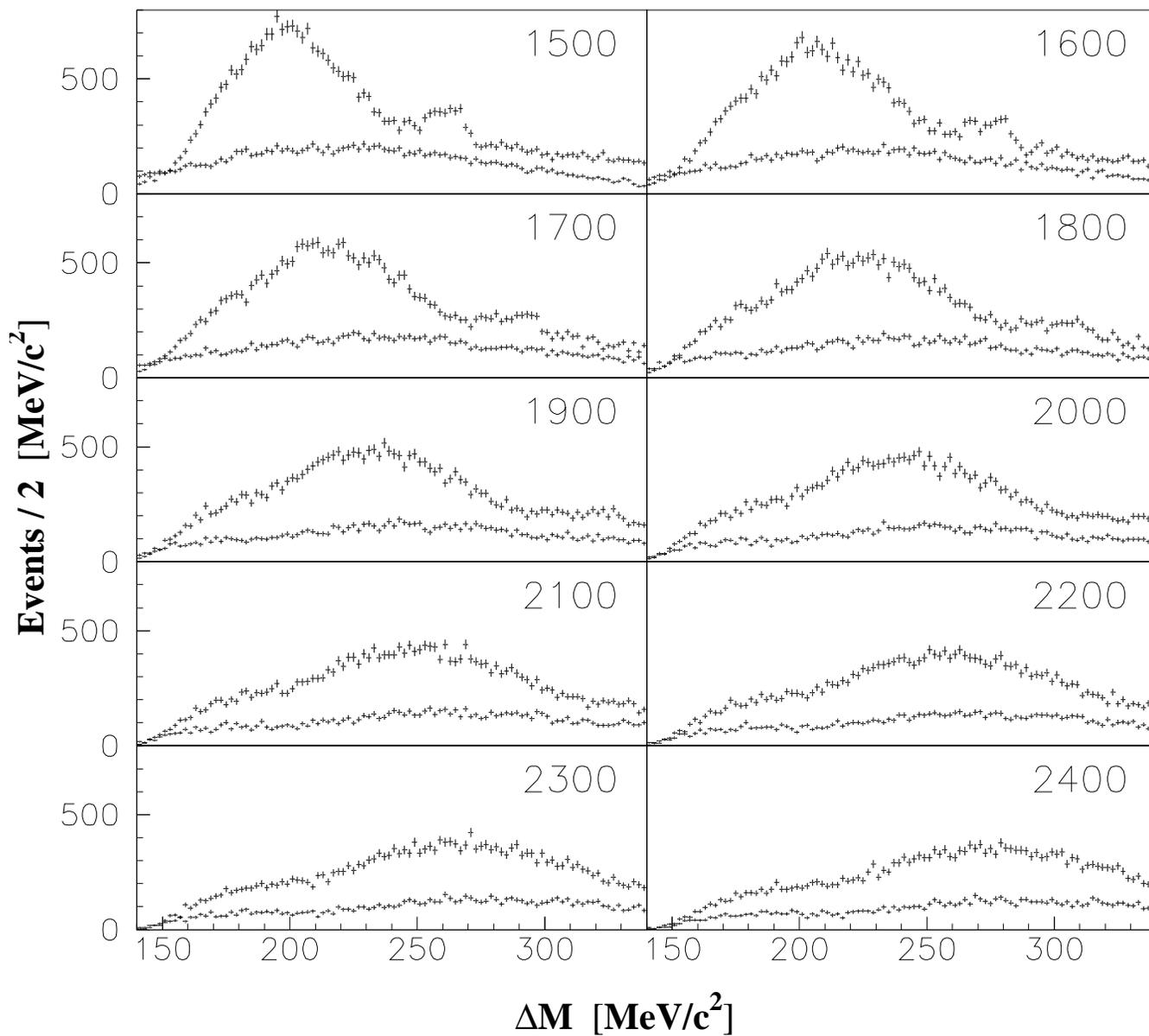

Fig. 3

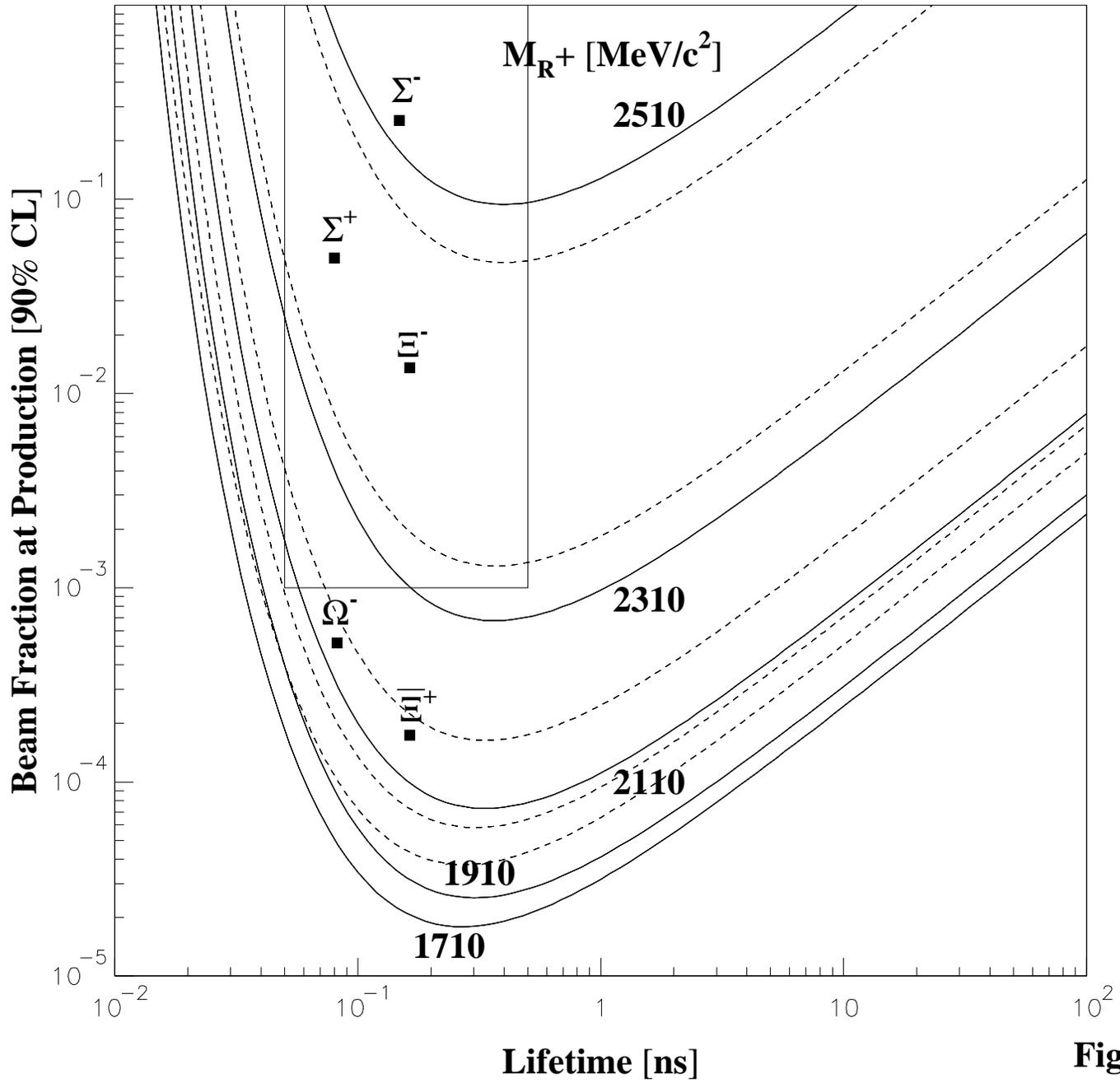

Fig. 4

# A Search for Light Super-Symmetric Baryons

I.F.Albuquerque[6(a)], N.F.Bondar[3], R.Carrigan Jr.[1], D.Chen[10(b)], P.S.Cooper[1], Dai Lisheng[2], A.S.Denisov[3], A.V.Dobrovolsky[3], T.Dubbs[5(c)], A.M.F.Endler[8], C.O.Escobar [6], M.Foucher[7(d)], V.L.Golovtsov[3], H.Gottschalk[1(e)], P.Gouffon[6(f)], V.T.Grachev[3], A.V.Khanzadeev[3], M.A.Kubantsev[4(g)], N.P.Kuropatkin[3], J.Lach[1], J.Langland[5], Lang Pengfei[2], Li Chengze[2], Li Yunshan[2], M.Luksys[13], J.R.P.Mahon[6(fh)], E.McCliment[5], A.Morelos[1(i)], C.Newsom[5], M.C.Pommot Maia[9], V.M.Samsonov[3], V.A.Schegelsky[3], Shi Huanzhang[2], V.J.Smith[11], Tang Fukun[2], N.K.Terentyev[3], S.Timm[12(j)], I.I.Tkatch[3], L.N.Uvarov[3], A.A.Vorobyov[3], Yan Jie[2], Zhao Wenheng[2], Zheng Shuchen[2], Zhong Yuanyuan[2]

[(1)] Fermi National Accelerator Laboratory, Batavia, IL 60510
[(2)] Institute of High Energy Physics, Beijing, PRC
[(3)] St. Petersburg Nuclear Physics Institute, Gatchina, Russia
[(4)] Institute of Theoretical and Experimental Physics, Moscow, Russia
[(5)] University of Iowa, Iowa City, IA 52242
[(6)] Universidade de Sao Paulo, Sao Paulo, Brazil
[(7)] J.W. Gibbs Laboratory, Yale University, New Haven, CT 06511
[(8)] Centro Brasileiro de Pesquisas Fisicas, Rio de Janeiro, Brazil
[(9)] Conselho Nacional de Pesquisas CNPq, Rio de Janeiro, Brazil
[(10)] State University of New York at Albany, Albany, NY 12222
[(11)] H.H. Wills Physics Laboratory, University of Bristol, UK
[(12)] Carnegie Mellon University, Pittsburgh, PA 15213
[(13)] Universidade Federal da Paraiba, Paraiba, Brazil

(E761 Collaboration)
February 22, 1996

## Abstract

We have searched for the production and decay of light super-symmetric baryons produced in 800 GeV/c proton copper interactions in a charged hyperon beam experiment. We observe no evidence for the decays $R^+(uud\tilde{g}) \to S^\circ(uds\tilde{g}) \pi^+$ and $X^-(ssd\tilde{g}) \to S^\circ(uds\tilde{g}) \pi^-$ in the predicted parent mass and lifetime ranges of 1700-2500 MeV/c$^2$ and 50-500 ps. Production upper limits for $R^+$ at $x_F=0.47$, $P_t=1.4$ GeV/c$^2$ and $X^-$ at $x_F=0.48$, $P_t=0.65$ GeV/c$^2$ of less than $10^{-3}$ of all charged secondary particles produced are obtained for all but the highest masses and shortest lifetimes predicted.

PACS numbers: 13.85.Rm, 14.80.Ly

Recent theoretical work[1,2,3] has proposed a super-symmetric model with a light gluino [$\tilde{g}$] in the mass range 100-600 MeV/$c^2$. A direct consequence of this model is the prediction of a set of super-symmetric hadrons in the 1000-3000 MeV/$c^2$ mass range. We have searched for two weak decays modes between these hadrons in an attempt to either confirm this hypothesis or close the low mass window for super-symmetry.

Among the light hadrons predicted[1,2,3] are the super-symmetric partners of the proton, $R^+$(uud$\tilde{g}$); $\Lambda°$, $S°$(uds$\tilde{g}$) and $\Xi^-$, $X^-$(ssd$\tilde{g}$). The predicted mass and lifetime ranges for these states are 1700-2500 MeV/$c^2$ and 50-500 ps. The mass splitting predicted[3] between these super-baryons is $\Delta M = M_{R^+} - M_{S°} = 210\pm20$ MeV/$c^2$. The $S°$ is less massive than the $R^+$ in this model[3] due to the very strong attraction of the flavor singlet quarks in the $S°$. Only strangeness changing weak decays are allowed between these lowest mass super-baryons. The dominant decays are expected to be $R^+ \rightarrow S° \pi^+$ and $X^- \rightarrow S° \pi^-$ analogous to normal hyperon decays, $\Sigma^+ \rightarrow n \pi^+$ and $\Xi^- \rightarrow \Lambda° \pi^-$. The mass, lifetime and decay modes predicted for these super-baryons mimic the hyperons. Limits on the production rate of $R^+$ and $X^-$ as a function of mass and lifetime can be set in charged hyperon beam searches. We report such a search.

Fermilab experiment E761 was a high statistics study of hyperon radiative decays. We reconstructed[4] $48 \times 10^6$ $\Sigma^+ \rightarrow p \pi°$ and $38 \times 10^3$ $\Sigma^+ \rightarrow p \gamma$ decays from data taken in 1990. Unfortunately, this data set cannot be used for an $R^+$ search since the trigger required a photon in the final state. In our observation of the $\Xi^- \rightarrow \Sigma^- \gamma$ decay[5] we took negative beam data in a different configuration of the apparatus. This had good acceptance for low momentum charged secondaries but also required a photon trigger. In order to control the photon trigger rate for these negative beam data we added a scintillator positioned to veto the pions from the $\Sigma^- \rightarrow n\pi^-$ mode. Subsequently, we removed the photon spectrometer from the apparatus and took data requiring only a beam particle and a signal in the pion veto scintillator as a trigger in order to measure the yields[6] and production polarizations[7] of the negative hyperons using their hadronic decay modes.  These data permitted a search for X⁻ → S° π⁻. As part of this study we briefly returned to positive beam running with the same trigger.  These data are available to search for R⁺ → S° π⁺.  Together these data comprise ~1% of the data taken by E761.

The trigger required only an incoming charged parent particle and a charged daughter with less than 40% of the incoming parent's momentum. The daughter is assumed to be a π±.  The neutral daughter is assigned the mass of the S°. Using momentum conservation to measure the momentum of the unseen neutral daughter we reconstruct the mass of the parent.  We search for a bump in the mass spectrum of the reconstructed parent particle as a function of the assumed S° mass.  After corrections for acceptance and decay losses we can determine upper limits on the fraction of R⁺ and X⁻ produced within our hyperon beam's acceptance as a function of the assumed parent mass and lifetime.  In the following text the parameters for the positive (negative) beam samples [R⁺(X⁻) decays] are shown as here.

The E761 charged hyperon beam was located in the Proton Center beam line at Fermilab.  800 GeV/c protons interacted in a 0.5x2.0 mm² x150 mm long copper target.  A magnet and beam channel downstream of the target selected positive (negative) secondaries with <P>=378 (382) GeV/c at a production angle near 3.7 (1.7) mrad.  The production variables of the hyperon beam were $x_F$=0.47 (0.48), $P_t$=1.4 (0.65) GeV/c with acceptances (FWHM) in these quantities of $\Delta x_F$=0.04  and $\Delta P_t$=0.4 GeV/c. 5.9(1.8) x10¹³ protons were incident on the production target for these data.  About half of these interacted in the target.  173 (169) x10⁶ secondaries were observed at the channel exit of the hyperon beam.  The larger secondary yield in the negative beam was due to the smaller production $P_t$.  3.45 (4.55) x10⁶ triggers were recorded with a livetime of 47 (35)%.

The standard E761 analysis and cuts are applied to these data[4,5,6]. Events are selected which have well reconstructed parent and daughter charged tracks with at least a 100 μrad decay angle [θ] between them, a vertex position in a 12 m long decay region beginning 13.6 m from the hyperon production target and a daughter to parent momentum ratio r<0.4. The details of the beam, apparatus and analysis are described elsewhere[4,5,6].

Figure 1 shows a θ-r scatter plot of the positive data. Superimposed are the two body kinematics curves for the known hyperon decays in this region and a set of curves for the decay $R^+ \rightarrow S^° \pi^+$ parametric in the $S^°$ mass and assuming the predicted value for ΔM. A large $\Sigma^+ \rightarrow n \pi^+$ component ($112 \times 10^3$ events) is clearly seen. The acceptance of the trigger drops rapidly below r=0.15 ($P_{\pi^+} \sim 50$ GeV/c) due to low momentum daughter tracks missing the trigger counter. Events remain at lower momenta due to interactions which fire the trigger counter even when the daughter misses. Our acceptance to the $R^+ \rightarrow S^° \pi^+$ decay is in the θ-r region, θ<3 mrad, 0.15<r<0.20.

In order to illustrate the sensitivity of our apparatus to two body decays we show $\Sigma^+ \rightarrow n \pi^+$ and $\bar{\Xi}^+ \rightarrow \bar{\Lambda}^° \pi^+$ events in these data. The results are shown as reconstructed parent mass plots in Fig. 2. In addition to the $\Sigma^+$ events (Fig. 2a) which were already obvious on the θ-r plot, we observe 1221 ± 59 $\bar{\Xi}^+$ events (Fig. 2b full curve) with a mass resolution [σ] of 1.9 MeV/c$^2$. Since the predictions of the model don't tell us the mass of either parent or daughter super-baryon we must be concerned with the dependence of our parent mass resolution when the incorrect daughter mass is assumed. We test this using the $\bar{\Xi}^+$ events. The dashed curve and fit in Fig. 2b are the same events analyzed using $M_{\Lambda^°}$ shifted 50 MeV/c$^2$ above the accepted value. The $\bar{\Xi}^+$ peak shifts as expected and the peak width increases from 1.9 to 2.7 MeV/c$^2$.

To search for the $R^+$ ($X^-$) $\rightarrow S^° \pi^{+(-)}$ decay modes we plot the reconstructed parent mass minus assumed $S^°$ mass, ΔM, for 10 different values of the $S^°$ mass spaced by 100 MeV/c$^2$ in Fig. 3. The $S^°$ would be reconstructed within 50 MeV/c$^2$ of its actual mass on one of these plots. These events are selected to have a reconstructed $\Sigma^{+(-)}$ mass below the $\Sigma^{+(-)} \rightarrow n \pi^{+(-)}$ mass band [M($\Sigma^{+(-)}$)<1179 (1185) MeV/c$^2$], be inconsistent with the $\Xi^{+(-)} \rightarrow \Lambda^° \pi^{+(-)}$ hypothesis (|M(Ξ)-1321| < 3 (6) MeV/c$^2$) and have a decay angle greater than 100 μrad. These cuts are shown as the heavy

curves on Fig. 1. This sample of 14127 positive events is shown on the insert of Fig. 1. The negative sample (not shown) contains 34345 events. The plots of Figs. 2 and 3 are in 2 MeV/$c^2$ mass bins which is the typical peak width [$\sigma$] we expect for any signal. Any single bin effect is too narrow to be consistent with our resolution and therefore must be a statistical fluctuation. There are 2000 data points displayed in Fig. 3 where both positive and negative data are shown on each plot. The number of >3$\sigma$ single bin fluctuations is consistent with expectations. There are no apparent signals visible in these data. The peaks in Fig. 2b demonstrate how a ~1000 event signal would appear in our apparatus.

We measure the fraction of all charged particles produced in the acceptance of the secondary beam which are of a particular particle type. To quantify upper limits to the production fraction of $R^+(X^-)$ we choose the predicted $R^+(X^-)$ $S°$ mass difference of $\Delta M=210$ MeV/$c^2$ and evaluate the 90% CL upper limit to the amplitude of a Gaussian with mean fixed at $\Delta M$ and width [$\sigma$] fixed at 2 MeV/$c^2$ in 5 of the 10 plots of Fig. 3. We take this limit to be twice the statistical error ($\sigma$~25 (50) events) to avoid the statistical fluctuations in the particular few bins near $\Delta M=210$ MeV/$c^2$ in each mass plot. Similar upper limits would apply to the case of a different mass splitting $\Delta M$. The upper limit scales as the square root of the number of events at the corresponding points in Fig. 3.

We use a simple Monte Carlo apparatus simulation to determine the mass resolution and acceptance for these decays. It reproduces the mass resolutions obtained from the known hyperon decays. The mass resolution for $R^+$ decays is 1.65 MeV/$c^2$ independent of $R^+$ mass, nearly the same value as observed for $\bar{\Xi}^+$ decays. This simulation does not take interactions into account and cannot reproduce the triggers below r=0.15. Given that we have no information on the interaction properties of super-baryons this is the safe course in determining an upper limit to production. After correction for branching ratios and the fraction of produced parents which decay in the decay region (decay factor) the fraction of particles of each type at production are shown in Table 1. The acceptances and decay factors are determined at the accepted lifetime of the $\Sigma^{+8}$ and the $R^+(X^-)$ branching ratio is assumed to be unity. These upper limits are shown as a function of the R+ (X-) mass and lifetime in Fig. 4. The lifetime dependence of the upper limit is given by the decay factor; $UL(\tau) = UL(\tau_{\Sigma+})D(\tau_{\Sigma+})/D(\tau)$ with:

$$D(\tau) = \exp(-L_p/(c\tau P/M)) [1 - \exp(-L_d/(c\tau P/M))]$$

where $L_p$ is the distance from production to the start of the decay region (13.6 m), $L_d$ is the length of the decay region (12.0 m), P is the average parent momentum [378 (382) GeV/c], M is the parent mass and $\tau$ is the parent lifetime.

Table I. Decay mode production fractions.

| Mode | acceptance | events [<90% CL] | decay factor | production fraction |
|---|---|---|---|---|
| $\Sigma^+ \to n\, \pi^+$ | 41% | 112200 | 0.134 | 5.0E-2 |
| $\bar{\Xi}^+ \to \Lambda^\circ\, \pi^+$ | 38% | 1221 | 0.219 | 1.7E-4 |
| $R^+[1710] \to S^\circ[1500]\, \pi^+$ | 20% | <62 | 0.070 | <5.1E-5 |
| $R^+[1910] \to S^\circ[1700]\, \pi^+$ | 13% | <53 | 0.054 | <9.1E-5 |
| $R^+[2110] \to S^\circ[1900]\, \pi^+$ | 4.0% | <47 | 0.041 | <3.4E-4 |
| $R^+[2310] \to S^\circ[2100]\, \pi^+$ | 0.43% | <45 | 0.031 | <4.0E-3 |
| $R^+[2510] \to S^\circ[2300]\, \pi^+$ | 1E-06 | <35 | 0.029 | <7.3E-1 |
| $\Sigma^- \to n\, \pi^-$ | 41% | 805770 | 0.220 | 2.5E-1 |
| $\Xi^- \to \Lambda^\circ\, \pi^-$ | 38% | 81946 | 0.219 | 1.4E-2 |
| $\Omega^- \to \Lambda^\circ\, K^-$ | 44% | 889 | 0.079 | 5.2E-4 |
| $X^-[1710] \to S^\circ[1500]\, \pi^-$ | 20% | <114 | 0.072 | <1.1E-4 |
| $X^-[1910] \to S^\circ[1700]\, \pi^-$ | 13% | <108 | 0.055 | <2.1E-4 |
| $X^-[2110] \to S^\circ[1900]\, \pi^-$ | 4.0% | <94 | 0.042 | <7.7E-4 |
| $X^-[2310] \to S^\circ[2100]\, \pi^-$ | 0.43% | <78 | 0.032 | <7.9E-3 |
| $X^-[2510] \to S^\circ[2300]\, \pi^-$ | 1E-06 | <66 | 0.024 | <3.7E-1 |

The total production cross section in 800 GeV/$c^2$ pN interactions for >3 GeV/$c^2$ gluinos has been calculated[9]. We extrapolate this calculation to the 100-600 MeV/$c^2$ gluino mass range and assume that gluinos hadronize into baryons 10% of the time in analogy with the strange quark.

The resulting production cross section for super-baryons is in the range 30-80 μb or (1-3) x$10^{-3}$ of the inelastic cross section. The extrapolation of this perturbative QCD calculation to such low masses is unreliable but provides an order of magnitude estimate for the super-baryon production fraction.

Qualitatively, if the gluino is as light as the s quark (current quark mass ~150 MeV/$c^2$) then it should be no more difficult to produce an $R^+$ than a $\Sigma^+$ hyperon. We clearly rule this case out. Charmed hadrons are produced with production fractions in the $10^{-3}$-$10^{-4}$ range. Since the charmed quark (~1500 MeV/$c^2$) is 2.5 times heavier than the heaviest gluino allowed by the model[1,2,3] one would expect to observe $R^+$ and $X^-$ production with beam fractions above $10^{-3}$. We rule this out for all but the highest mass and shortest lifetime super-baryons allowed by the model.

Neither of these production fraction estimates is a calculation of the super-baryon production cross section. They do indicate a level (~$10^{-3}$) at which upper limits become meaningful in constraining models of this type. We observe no evidence for the production and subsequent pionic decay of super-baryons with upper limits below $10^{-3}$ of the total cross section over much of the allowed parameter space of this model.

We acknowledge helpful conversations with several of our Fermilab colleagues; Jeff Appel, Estia Eichten and Chris Quigg. We wish to thank the staffs of Fermilab and the Petersburg Nuclear Physics Institute for their assistance. This work is supported in part by the U.S. Department of Energy under contracts DE-AC02-76CH03000, DE-AC02-76ER03075, DE-FG02-91ER40631, DE-FG02-91ER40664, DE-FG02-91ER40682, the Russian Academy of Sciences, and the UK Science and Engineering Research Council.

____________________

[a] Supported by FAPESP, Brazil, Now at Rutgers University, Piscataway, NJ
[b] Now At Fermilab, Batavia, IL 60510

(c) Present address, Department of Physics, University of California (Santa Cruz), Santa Cruz, CA 95046.
(d) Present address, Department of Physics, University of Maryland, College Park, MD 20742
(e) Present address, *COPPE,* Universidade Federal do Rio de Janeiro, Brazil
(f) Partially supported by FAPESP and CNPq, Brazil
(g) Partially supported by the International Science Foundation
(h) Now at the Universidade Estadual do Rio de Janeiro, Brazil.
(i) Partially supported by CONACyT, México.
   Present address, Instituto de Física, University Autónoma de San Luis Potosí, San Luis Potosí, S.L.P. 78240 México.
(j) Present address, Department of Physics, SUNY Albany, Albany NY 12222

(c) Present address, Department of Physics, University of California (Santa Cruz), Santa Cruz, CA 95046.
(d) Present address, Department of Physics, University of Maryland, College Park, MD 20742
(e) Present address, *COPPE,* Universidade Federal do Rio de Janeiro, Brazil
(f) Partially supported by FAPESP and CNPq, Brazil
(g) Partially supported by the International Science Foundation
(h) Now at the Universidade Estadual do Rio de Janeiro, Brazil.
(i) Partially supported by CONACyT, México.
   Present address, Instituto de Física, University Autónoma de San Luis Potosí, San Luis Potosí, S.L.P. 78240 México.
(j) Present address, Department of Physics, SUNY Albany, Albany NY 12222

## References

1. G. Farrar, Phys. Rev. D51, 3904 (1995).
2. G. Farrar, RU-95-25 July 11, 1995, hep-ph/9508291.
3. G. Farrar, RU-95-26 July 11, 1995, hep-ph/9508292, submitted to Phys. Rev. Letters.
4. M. Foucher *et al.,* Phys. Rev. Lett. 68, 3004 (1992).
5. T. Dubbs *et al.,* Phys. Rev. Lett. 72, 808 (1994).
6. Jerry Langland, Ph.D. Thesis, University of Iowa, 1995. (unpublished).
7. Steven Timm, Ph.D. Thesis, Carnegie Mellon University, 1995. (unpublished).
8. Particle Data Group, Phys.Rev. D50, 1173 (1994).
9. S. Dawson, E. Eichten and C. Quigg Phys. Rev. D31, 1581 (1985) and private communication E. Eichten & C. Quigg.

# Figure Captions

1. $\theta$-r plot for the positive event sample. The solid curves shown are the expected locations of two body decay modes of 378 GeV/c parents in these variables. The modes shown are (from right to left) $\bar{\Omega}^+ \to \bar{\Lambda}^\circ K^+$, $\Sigma^+ \to n\pi^+$, $\bar{\Xi}^+ \to \bar{\Lambda}^\circ \pi^+$, $R^+ \to S^\circ \pi^+$ with $M_{S^\circ}$ = 1500, 1700, 1900, 2100, 2300 MeV/c$^2$; $\Delta M$=210 MeV/c$^2$. The heavy curves show the cuts for the $R^+$ search. The insert shows the same distribution after the cuts have been imposed (14127 events). The histogram displays the r distribution for these events.

2. Reconstructed parent mass plots for the known hyperon decays in the data sample. The fits shown are Gaussian plus linear backgrounds. 2(a) The $\Sigma^+ \to n\pi^+$ sample. 2(b) The $\bar{\Xi}^+ \to \bar{\Lambda}^\circ \pi^+$ sample. The ordinate is the reconstructed parent mass minus the $\Lambda^\circ$ mass. The dashed curve and fit are the same events analyzed with an assumed $\Lambda^\circ$ mass shifted 50 MeV/c$^2$ above the accepted value.

3. $R^+$ and $X^-$ mass plots for assumed $S^\circ$ masses from 1500-2400 MeV/c$^2$ as a function of $\Delta M$, the $R^+$ ($X^-$), $S^\circ$ mass difference. The $X^-$ data are the upper curves on each plot. The broad structure near $\Delta M$=275 MeV/c$^2$ in the $X^-$ data at low $S^\circ$ mass is an artifact of the $\Xi^- \to \Lambda^\circ \pi^-$ cut.

4. Production fraction 90%CL upper limits for $R^+$(solid) and $X^-$(dashed) as a function of parent mass and lifetime. The observed hyperon production fractions are also shown. The box shown is the predicted lifetime range and our estimate of 10$^{-3}$ for the scale at which such particles might be produced.